\documentclass[aps,prd,onecolumn,groupedaddress,showpacs,nofootinbib,amssymb]{revtex4}
\usepackage[dvips]{graphicx}
\usepackage{amssymb}
\usepackage{amsmath}
\usepackage{graphicx,color}
\usepackage{amsfonts}
\usepackage{bm}
\usepackage{cancel}
\usepackage{comment}
\usepackage{hyperref}

\newcommand{\e}{\mathrm{e}}

\allowdisplaybreaks[4]

\begin{document}

\tolerance=5000

\title{Micro-canonical and canonical description for generalised entropy}

\author{Shin'ichi~Nojiri$^{1,2}$\,\thanks{nojiri@gravity.phys.nagoya-u.ac.jp},
Sergei~D.~Odintsov$^{3,4}$\,\thanks{odintsov@ieec.uab.es}}
\affiliation{
$^{1)}$ Department of Physics, Nagoya University,
Nagoya 464-8602, Japan \\
$^{2)}$ Kobayashi-Maskawa Institute for the Origin of Particles
and the Universe, Nagoya University, Nagoya 464-8602, Japan \\
$^{3)}$ ICREA, Passeig Luis Companys, 23, 08010 Barcelona, Spain\\
$^{4)}$ Institute of Space Sciences (ICE, CSIC) C. Can Magrans s/n, 08193 Barcelona, Spain}

\begin{abstract}
Few parameters dependent generalised entropy includes Tsallis entropy, R{\'e}nyi entropy, 
Sharma-Mittal entropy, Barrow entropy, Kaniadakis entropy, etc as particular representatives. 
Its relation to physical systems is not always clear. 
In this paper, we propose the microscopic thermodynamic description for an arbitrary generalised entropy in terms of the particle system. 
It is shown that the change in the volume of the phase space of the particle system in the micro-canonical description 
or the difference in the integration measure in the phase space in the canonical description may lead to generalised entropy. 
Our consideration may help us understand the structure of quantum gravity. 

\end{abstract}

\maketitle

\section{Introduction}\label{Sec1}

Recently generalised entropy which depends on several parameters has been proposed in Refs.~\cite{Nojiri:2022aof, Nojiri:2022dkr} 
(specifically, three, four, five, and six parameters dependent generalised entropy was constructed in these works). 
It generalizes all known previous entropies like Tsallis entropy~\cite{tsallis} (see also \cite{Ren:2020djc, Nojiri:2019skr}), 
R{\'e}nyi entropy~(\cite{renyi}, 
Sharma-Mittal entropy, Barrow entropy~\cite{Barrow:2020tzx}, 
Kaniadakis entropy~\cite{Kaniadakis:2005zk,Drepanou:2021jiv}, 
Loop Quantum Gravity entropy~\cite{Majhi:2017zao}, etc. 
Such entropies have been proposed for the description of different physical, statistical, and information systems.

Some motivation for the study of different entropies is also related to cosmology and gravity. 
Starting with the discovery of Hawking radiation, the thermodynamical structure of black holes has been investigated and it has been found that 
the entropy of a black hole is proportional to the horizon area. 
The black hole entropy is called Bekenstein-Hawking entropy \cite{Bekenstein:1973ur, Hawking:1975vcx}. 
This suggests that gravity should be deeply related to thermodynamics. 
Motivated by this, the thermodynamics of the Hubble horizon has been applied to 
cosmology.
It has been shown that the various expressions of entropy may lead to different holographic cosmologies~\cite{Nojiri:2021iko, Nojiri:2021jxf} and holographic dark energy 
models~\cite{Li:2004rb,Li:2011sd,Nojiri:2005pu,Gong:2004cb,Khurshudyan:2016gmb,Landim:2015hqa,
Gao:2007ep,Li:2008zq}. 
Such a holographic approach is valid even for the understanding of inflation at the early 
universe~\cite{Nojiri:2019kkp}, which allows the unified understanding of the 
dark energy and inflation by holographic cosmology. 

The microscopic understanding of generalised entropy could be helpful to clarify the structure of quantum gravity. 
In this paper, we consider how various entropies can appear from microscopic particle descriptions of the thermodynamical system. 
 
In the next section, we start with the micro-canonical description and consider how the corresponding particle system may give generalised entropy. 
In Section~\ref{Sec3}, by using the canonical approach, we consider the origin of generalised entropy again from the corresponding particle system. 
A summary is given in the last section. 

\section{Micro-canonical description}\label{Sec2}

In this section, based on the micro-canonical approach of thermodynamics, we consider the isolated system with 
fixed energy $E$ and clarify how various generalised entropies appear. 

An expression of the standard Gibbs entropy is given by
\begin{align}
\label{S1}
S (E) = - k \sum_{i=1}^{W (E)} P_i (E) \ln P_i (E) \, .
\end{align}
Here $k$ is the Boltzmann constant. 
We assume that there are ${W (E)}$ states with a fixed energy $E$ and $P_i(E) $ is a probability realizing the $i$-th state with the energy $E$ and therefore 
\begin{align}
\label{S2}
\sum_{i=1}^{W (E)} P_i (E) = 1 \, .
\end{align}
In \cite{tsallis}, Tsallis proposed a generalization of the entropy with a parameter $q$, 
\begin{align}
\label{S3}
S_q (E) \equiv k\frac{ 1 - \sum_{i=1}^{W (E)} \left( P_i (E) \right)^q}{q-1} 
= k \sum_{i=1}^{W (E)} \frac{P_i (E) \left( 1 - \left(P_i (E) \right)^{q-1} \right)}{q-1} \, .
\end{align}
In the limit of $q\to 1$, $S_q (E) $ reduces to the standard expression in (\ref{S1}). 

As a further generalization, we may consider the following expression, 
\begin{align}
\label{S4}
S (E) = k \sum_{i=1}^{W (E)} s_i \left( P_i (E) \right) \, .
\end{align}
When we regard $S (E)$ as a function of $P_i$, we may consider the maximum of $S (E)$ in (\ref{S4}) under the condition (\ref{S2}), 
which corresponds to the thermal equilibrium. 
For this purpose, we introduce the Lagrange multiplier constant $\lambda$ and consider the stationary point of the following 
quantity, 
\begin{align}
\label{S5}
L = k \left\{ \sum_{i=1}^{W (E)} s_i \left( P_i (E) \right) - \lambda \left( \sum_{i=1}^{W (E)} P_i (E) - 1 \right) \right\} \, .
\end{align}
The stationary point is given by the following equation, 
\begin{align}
\label{S6}
0= \frac{ds_i \left( P_i (E) \right)}{d P_i (E) } - \lambda \, ,
\end{align}
which can be solved with respect to $P_i (E) $ as $P_i (E) = P_i \left( E, \lambda \right)$. 
The constant $\lambda$ can be determined by the condition (\ref{S2}), 
\begin{align}
\label{S7}
\sum_{i=1}^{W (E)} P_i \left(E, \lambda\right) = 1 \, .
\end{align}
In the case that $s_i( \xi )$ does not depend on $i$, $s_i( \xi )=s(\xi)$, because $\lambda$ is a constant, $P_i (E)$ which 
is a solution of (\ref{S6}), does not depend on $i$, either, and Eq.(\ref{S7}) tells 
\begin{align}
\label{S8}
P_i (E) =P(E, \lambda)=\frac{1}{{W (E)}}\, ,
\end{align}
which is the equation determining $\lambda$. 
Therefore the entropy in the thermal equilibrium is given by 
\begin{align}
\label{S9}
S (E) = k {W (E)} s \left( \frac{1}{{W (E)}} \right) \, .
\end{align}
Thus we obtain an expression for the generalised entropy in the thermal equilibrium. 
In the micro-canonical approach, the temperature $T$ is defined by 
\begin{align}
\label{T}
\frac{1}{T} \equiv \frac{dS(E)}{dE}\, .
\end{align}

Instead of considering the discrete states, we may consider the phase space of $N$ particles $\left(q^i, p_i \right)$ $\left( i=1,2,\cdots, N \right)$, 
which corresponds to $W(E)\to \infty$ limit. 
Then the equations corresponding to (\ref{S4}) and (\ref{S5}) have the following forms, 
\begin{align}
\label{S10}
S=&\, k \int_E \prod_{i=1}^N \left( \frac{dq^i dp_i}{\hbar} \right) s \left( q^k, p_k, P\left( q^k, p_k, E \right) \right)\, , \\
\label{S11}
L=&\, k \left\{ \int_E \prod_{i=1}^N \left( \frac{dq^i dp_i}{\hbar} \right) s \left( q^k, p_k, P\left( q^k, p_k, E \right) \right) 
 - \lambda \left( \int_E \prod_{i=1}^N \left( \frac{dq^i dp_i}{\hbar} \right) P\left( q^k, p_k, E \right) - 1 \right) \right\} \, .
\end{align}
Note that $s$ can explicitly depend on $q^i$ and $p_i$, in general 
and $\int_E \prod_{i=1}^N \left( \frac{dq^i dp_i}{\hbar} \right) \cdots $ is the integration of the phase space for fixed energy $E$. 

The function $s$ may include a finite or infinite number of parameters, $\left\{\alpha_n\right\}$, $n=1,2,\cdots$ as 
$s=s \left( \left\{\alpha_n\right\}; q^k, p_k, P\left( q^k, p_k, E \right) \right)$, 
In the limit of the parameters, $\left\{\alpha_n\right\}$, $s$ may reduce to that in the Gibbs entropy (\ref{S1}). 
As in the case of the black hole entropy in \cite{Nojiri:2022ljp}, we may impose the following conditions for $s$ with the parameters $\left\{\alpha_n\right\}$:
\begin{enumerate}
\item Generalised third law: The generalised entropy vanishes when temperature $T$ vanishes as in the case of the Gibbs entropy (\ref{S1}). 
Note, however, Bekenstein-Hawking entropy $S_\mathrm{BH}$ for black hole diverges when Hawking temperature $T_\mathrm{H}$ vanishes and 
$S_\mathrm{BH}$ vanishes when $T_\mathrm{H}\to \infty$. 
\item Monotonicity: The generalised entropy is a monotonically increasing function of Gibbs entropy (\ref{S1}). 
\item Positivity: The generalised entropy should be positive, as the number of states is greater than unity.
\item Gibbs entropy limit: The generalised entropy reduces to the Gibbs entropy (\ref{S1}) in an appropriate limit of the parameters $\left\{\alpha_n\right\}$. 
\end{enumerate}
In the standard thermodynamics, we also need to impose the zeroth law, which tells
\begin{itemize}
\item When two systems denoted by $A$ and $B$ are in thermal equilibrium with a third system denoted by $C$, the system $A$ is also in equilibrium with the system $B$. 
\end{itemize}
In the case of non-extensive entropies like the Tsallis entropy, the above zeroth law does not hold \cite{Nauenberg:2002azf}. 
Therefore the generalised entropies which we consider in this paper do not always satisfy the zeroth law. 

In the thermal equilibrium, we find 
\begin{align}
\label{S12}
0 = \frac{d s \left( q^k, p_k, P\left( q^k, p_k, E \right) \right)}{d P\left( q^k, p_k, E \right)} - \lambda \, ,
\end{align}
which can be solved again with respect to $P\left( q^k, p_k, E \right)$ as 
\begin{align}
\label{S13}
P\left( q^k, p_k, E \right)=P\left( q^k, p_k, E,\lambda \right) \, ,
\end{align}
and $\lambda$ is determined by the equation
\begin{align}
\label{S14}
\int_E \prod_{i=1}^N \left( \frac{dq^i dp_i}{\hbar} \right) P\left( q^k, p_k, E, \lambda \right) =1 \, .
\end{align}
In the case that $s$ does not explicitly depend on $q^i$ and $p_i$, $s=s \left( P\left( q^k, p_k, E \right) \right)$, 
Eq.~(\ref{S12}) shows that $P$ does not explicitly depend on $q^i$ and $p_i$, either, 
\begin{align}
\label{S15}
P = P(E,\lambda)\, ,
\end{align}
and by using (\ref{S14}), we find 
\begin{align}
\label{S15}
P = P(E,\lambda)= \frac{1}{V_\mathrm{phase}}\, , \quad V_\mathrm{phase} \equiv \int_E \prod_{i=1}^N \left( \frac{dq^i dp_i}{\hbar} \right) \, ,
\end{align}
and 
\begin{align}
\label{S16}
S= k V_\mathrm{phase} s \left( \frac{1}{V_\mathrm{phase}} \right) \, .
\end{align}
Here $V_\mathrm{phase}$ is the volume of the phase space, which can be finite because we are considering the fixed energy $E$.
By the choice of $s$, one can obtain several kinds of entropy. 

As a simple example, we consider one non-relativistic particle with mass $m$ moving on the two-dimensional space whose area is $A$. 
Because 
\begin{align}
\label{S17}
E=\frac{{p_x}^2 + {p_y}^2}{2m}\, , 
\end{align}
the volume of the momentum space is the area of a two-dimensional sphere with the radius $\sqrt{2m E}$, that is, $4\pi \left( \sqrt{2m E} \right)^2 = 8\pi m E$. 
Therefore we find 
\begin{align}
\label{S18}
V_\mathrm{phase}= \frac{8\pi m E A}{\hbar^3}\, .
\end{align}
If we choose $s(\xi)=-x\ln \xi$, we obtain the standard expression of the Gibbs entropy, which we now denote $S_0$, 
\begin{align}
\label{S19}
S_0 = k \ln V_\mathrm{phase} = k \ln \left( \frac{8\pi m E A}{\hbar^3} \right)\, .
\end{align}
On the other hand, if $s(\xi)$ is given by 
\begin{align}
\label{S20}
s(\xi) = \frac\xi{\gamma} \left[ \left( 1 - \frac{\alpha}{\beta} \ln \xi \right)^\beta - 1 \right] \, ,
\end{align}
with positive dimensionless parameters $ \left( \alpha , \beta, \gamma \right)$, 
or $\left\{\alpha_n\right\}=\left\{ \alpha, \beta, \gamma \right\}$, 
we obtain an expression similar to the three-parameter entropy in \cite{Nojiri:2022aof}, 
\begin{align}
\label{S21}
S_\mathrm{G} \left( \alpha, \beta, \gamma \right)
= \frac{ k}{\gamma} \left[ \left( 1 + \frac{\alpha}{\beta { k}} S_0 
\right)^\beta - 1 \right] \, .
\end{align}
{ 
Here $S_0$ is given in (\ref{S19}). 
Eq.~(\ref{S21}) is obtained by substituting (\ref{S18}) into (\ref{S16}) with (\ref{S20}) and by using (\ref{S19}). 
} 
On the other hand, in the case, that $s(\xi)$ is given by 
\begin{align}
\label{S22}
s(\xi) = \frac\xi{\gamma}\left[\left(1 - \frac{\alpha_+}{\beta} \ln \xi \right)^{\beta} 
 - \left(1 - \frac{\alpha_-}{\beta} \ln \xi \right)^{-\beta}\right] \,,
\end{align}
we obtain an expression corresponding to a four-parameters generalised entropy proposed in \cite{Nojiri:2022dkr}, 
\begin{align}
\label{S23}
S_\mathrm{G} \left( \alpha_+ , \alpha_- , \beta , \gamma \right)
= \frac{ k}{\gamma}\left[\left(1 + \frac{\alpha_+}{\beta{ k}} S_0\right)^{\beta} 
 - \left(1 + \frac{\alpha_-}{\beta{ k}} S_0 \right)^{-\beta}\right] \, .
\end{align}
Here $\left\{\alpha_n\right\}=\left\{ \alpha_\pm, \beta, \gamma \right\}$ { and $\alpha_\pm$ are dimensionless parameters}.  
We can find $s(\xi)$ corresponding to other versions of generalised entropy. 

Hence, starting from the micro-canonical approach of thermodynamics, where we consider the isolated system with 
fixed energy $E$, we derived the various versions of entropy. 
The entropy is specified by the function $s_i( \xi )$ in (\ref{S10}).

%%%%%%%%%%%%%%%%%%%%%%%%%%%%%%%%%%%%%%%%%%%%%

We may speculate on the origin of the generalised entropy (\ref{S4}). 
In the case of non-additive systems, such as gravitational or electromagnetic ones, 
the standard Gibbs additive entropy (\ref{S1}) should be generalised to the non-extensive Tsallis entropy~\cite{tsallis}. 
The non-extensive entropy suggests that the numbers of the states show the running behavior by the change of the energy scale. 
In quantum field theory, such a running behavior usually appears via the renormalization group. 
Since the entropy corresponds to the physical degrees of freedom of a system, the renormalization group 
of a quantum theory implies that the degrees of freedom depend on the energy scale. 
In the low-energy regime, massive modes decouple, and therefore the degrees of freedom decrease. 
In the case of gravity, if the space-time fluctuations become large in the ultraviolet regime, the
degrees of freedom may increase. 
On the other hand, if gravity becomes topological, the degrees of freedom decrease. 
The latter situation is consistent with holography. 
Based on such a consideration, Barrow has proposed one particular example of Tsallis-like entropy~\cite{Barrow:2020tzx}. 
This may suggest that the generalised entropy might also appear by reflecting the quantum structure of gravity.

\section{Canonical description}\label{Sec3}

In this section, we use the canonical approach of thermodynamics, where the system is in equilibrium with the heat bath with temperature $T$. 
We show that various versions of entropy may appear due to the difference in the integration measure in the phase space. 

Usually, the partition function for the statistical system of $N$ particles is defined by 
\begin{align}
\label{GS1}
Z({ \bar\beta}) = \int \prod_{i=1}^N \left( \frac{dq^i dp_i}{\hbar} \right) \e^{-{ \bar\beta} H \left( q^i, p_i \right)}\, .
\end{align}
Here $q^i$ and $p^i$ are the coordinate of the position and the momentum for the $i$-th particle 
{ $\bar\beta \equiv \frac{1}{kT}$}. 
The measure $ \prod_{i=1}^N \left( \frac{dq^i dp_i}{\hbar} \right)$ is chosen because it is invariant under the canonical transformation 
in classical mechanics. 
In quantum mechanics, however, the cartesian coordinates have a special meaning. 
As an example, we consider a free particle with mass $m$ in two spatial dimensions. 
If we use the cartesian coordinates $(x,y)$, the Hamiltonian is given by 
\begin{align}
\label{GS2}
H=\frac{{p_x}^2 + {p_y}^2}{2m} \, .
\end{align}
Here $p_x$ and $p_y$ are momenta conjugate to $x$ and $y$, respectively. 
To obtain the 
%''
Schr\"{o}dinger equation, we may replace the momenta with the partial derivatives, $p_x\to -i\hbar \partial_x$, $p_y\to -i\hbar \partial_y$, 
\begin{align}
\label{GS3}
H= - \frac{\hbar^2}{2m} \left( {\partial_x}^2 + {\partial_y}^2 \right) \, .
\end{align}
On the other hand, if we use the polar coordinate $(r,\phi)$ instead of the cartesian coordinates $(x,y)$, 
\begin{align}
\label{GS4}
x=r \cos\phi\, , \quad y=r\sin\phi\, .
\end{align}
we obtain the following Hamiltonian 
\begin{align}
\label{GS5}
H=\frac{1}{2m} \left( {p_r}^2 + \frac{{p_\phi}^2}{r^2} \right)\, .
\end{align}
Here $p_r$ and $p_\phi$ are momenta conjugate to $r$ and $\phi$, respectively. 
If we naively replace the momenta with the partial derivatives, $p_r\to -i\hbar \partial_r$, $p_\phi\to -i\hbar \partial_\phi$, 
we obtain 
\begin{align}
\label{GS6}
H= - \frac{\hbar^2}{2m} \left( {\partial_r}^2 + \frac{1}{r^2}{\partial_\phi}^2 \right) \, , 
\end{align}
which is different from the Hamiltonian in (\ref{GS3}). 
In fact, by using the polar coordinates, the Hamiltonian~(\ref{GS3}) can be rewritten as 
\begin{align}
\label{GS7}
H= - \frac{\hbar^2}{2m} \left( {\partial_r}^2 + \frac{1}{r}\partial_r + \frac{1}{r^2}{\partial_\phi}^2 \right) \, .
\end{align}
In classical mechanics, the coordinate transformation is given by the canonical transformation. 
Therefore the above result shows that quantum mechanics is not invariant under the canonical transformation. 
This may tell that we might not need to choose the measure by $\prod_{i=1}^N \left( \frac{dq^i dp_i}{\hbar} \right)$. 
The symmetry could give a guide to choosing the measure but the measure is not still uniquely determined. 

Instead of $\prod_{i=1}^N \left( \frac{dq^i dp_i}{\hbar} \right)$, we consider more general measure 
$\e^{-M\left(q^i,p_i\right)}\prod_{i=1}^N \left( \frac{dq^i dp_i}{\hbar} \right)$ and instead of (\ref{GS1}), we define the partition function 
\begin{align}
\label{GS8}
Z({ \bar\beta}) = \int \prod_{i=1}^N \left( \frac{dq^i dp_i}{\hbar} \right) \e^{-{ \bar\beta} H \left( q^i, p_i \right)- M\left(q^i,p_i\right)}\, .
\end{align}
When particles are confined on the box with edge length $L$, $M\left(q^i,p_i\right)$ could be given by 
\begin{align}
\label{GS9}
\e^{-M\left(q^i,p_i\right)} = \prod_{i=1}^N \theta \left( q^i \right) \theta \left( L - q^i \right) \, , \quad \mbox{or} \quad 
M = - \sum_{i=1}^N \left( \ln \theta \left( q^i \right) + \ln \theta \left( L - q^i \right) \right) \, .
\end{align}
Here $\theta(\xi)$ is the Heaviside step function defined by 
\begin{align}
\label{CS10}
\theta(\xi) = \left\{ \begin{array}{cc} 1 & \mbox{when}\ \xi \geq 0 \\
0 & \mbox{when}\ \xi<0 
\end{array} \right. \, .
\end{align}

As an example, we may consider the following model, 
\begin{align}
\label{CS11}
\left(q^i\right) = \left( x,y,z\right)\, , \quad 
H=\frac{1}{2m} \left( p_x^2 + p_y^2 + p_z^2 \right) \, , \quad 
\e^{-M}= { 4\pi R^2} \delta\left( R^2 - x^2 - y^2 -z^2 \right) \e^{-X\left( 4\pi \left( x^2 + y^2 + z^2 \right)\right)}\, .
\end{align}
Here $X$ is an adequate function. 
By performing the integrations, we find
\begin{align}
\label{CS12}
Z({ \bar\beta})= \frac{ 8\pi^2 R^3}{\hbar^3} \left( \frac{2m \pi}{{ \bar\beta}} \right)^\frac{3}{2} \e^{-X\left(4\pi R^2 \right)}\, ,
\end{align}
and therefore we obtain the free energy $F({ \bar\beta})$ given by 
\begin{align}
\label{CS13}
F({ \bar\beta}) = - \frac{1}{ \bar\beta} \ln Z({ \bar\beta}) 
= - \frac{1}{ \bar\beta} { \left( \ln \left( \frac{8\pi^2 R^3}{\hbar^3} \left( \frac{2m \pi}{\bar\beta} \right)^\frac{3}{2} \right)
 - X\left(4\pi R^2 \right) \right)} \, .
\end{align}
Then the thermodynamical energy $E\left({ \bar\beta} \right)$ is obtained as 
\begin{align}
\label{CS14}
E({ \bar\beta}) = F({ \bar\beta}) + { \bar\beta} \frac{\partial F({ \bar\beta})}{\partial { \bar\beta}} = \frac{3}{2{ \bar\beta}}\, ,
\end{align} 
and the entropy $S$ is 
\begin{align}
\label{CS15}
S={ k \bar\beta}\left( E-F \right) =  { k \left\{ \frac{3}{2} 
+ \ln \left( \frac{8\pi^2 R^3}{\hbar^3} \left( \frac{2m \pi}{\bar\beta} \right)^\frac{3}{2} \right) - X\left(4\pi R^2 \right) \right\} } \, .
\end{align}
If ${ X(\xi)=- \frac{c^3 \xi}{4\hbar G}}$ with Newton's gravitational constant $G$, the last term dominates for large $R$ and we obtain Bekenstein-Hawking entropy, 
\begin{align}
\label{CS16}
S\to S_\mathrm{BH}=\frac{{ k c^3} A}{4{ \hbar} G}\, , \quad A\equiv 4\pi R^2\, .
\end{align}
{ Here $c$ is the speed of light. }
On the other hand, if ${ X(\xi)=- \frac{c^3 A_0}{4\hbar G}\left( \frac{\xi}{A_0}\right)^\delta}$, we obtain Tsallis entropy, 
\begin{align}
\label{CS17}
S\to \frac{{ k c^3} A_0}{4{ \hbar} G}\left( \frac{A}{A_0}\right)^\delta\, .
\end{align}
The function $M\left(q^i,p_i\right)$, which gives the measure, will be determined by the properties of the physical system which we are considering 
but it is clear what kind of measure gives the corresponding kind of entropy. 

In the case of R{\'e}nyi entropy, which is defined as 
\begin{align}
\label{RS1}
S_\mathrm{R}=\frac{ k}{\alpha} \ln \left( 1 + { \frac{\alpha}{k}} S_\mathrm{BH} \right) \, ,
\end{align}
with a parameter $\alpha$, we find $X(\xi)=-\frac{1}{\alpha} \ln \left( 1 + \frac{\alpha \xi}{4G} \right)$. 

For the three-parameter entropy in \cite{Nojiri:2022aof} 
\begin{align}
\label{general6}
S_\mathrm{G} \left( \alpha, \beta, \gamma \right)
= \frac{ k}{\gamma} \left[ \left( 1 + \frac{\alpha}{\beta { k}} S_\mathrm{BH} 
\right)^\beta - 1 \right] \,,
\end{align}
with positive parameters $ \left( \alpha , \beta, \gamma \right)$, we obtain 
\begin{align}
\label{CS18}
X(\xi) = - \frac{1}{\gamma} \left[ \left( 1 + \frac{\alpha { c^3 \xi}}{4{ \hbar} G \beta} \right)^\beta - 1 \right] \,,
\end{align}
Further, a four-parameters generalised entropy proposed in \cite{Nojiri:2022dkr}, 
\begin{align}
\label{general1}
S_\mathrm{G} \left( \alpha_+ , \alpha_- , \beta , \gamma \right)
= \frac{ k}{\gamma}\left[\left(1 + \frac{\alpha_+}{\beta{ k}} S_\mathrm{BH}\right)^{\beta} 
 - \left(1 + \frac{\alpha_-}{\beta{ k}}S_\mathrm{BH}\right)^{-\beta}\right] \,,
\end{align}
with positive parameters $\left( \alpha_+ , \alpha_- , \beta , \gamma \right) $, $X(x)$ is given by 
\begin{align}
\label{CS19}
X(\xi) = - \frac{1}{\gamma}\left[\left(1 + \frac{\alpha_+ { c^3 \xi}}{4 { \hbar} \beta} G\right)^\beta 
 - \left(1 + \frac{\alpha_- { c^3 \xi}}{4 { \hbar} G \beta} \right)^{-\beta}\right] \,,
\end{align}
Thus we can always find the function $X(x)$ corresponding to the generalised entropy. 

The general measure can be related to the modification of the commutation relation 
$\left[ q^i, p_j \right]=i \hbar \delta^i_{\ j}$. 
First, we consider the case that $N=1$, that is, one pair of the canonical variables $(q,p)$, as follows, 
\begin{align}
\label{CS20}
\left[ q, p \right] = i\hbar \e^{M(q,p)}\, .
\end{align}
If we consider, however, the following new variable $Q$,
\begin{align}
\label{CS21}
Q=\int dq \e^{-M(q,p)} \, ,
\end{align}
we obtain the standard commutation relation, 
\begin{align}
\label{CS22}
\left[ Q, p \right] = i\hbar \, .
\end{align}
If we define the standard measure by using $Q$ and $p$, we obtain the previous generalised measure, 
\begin{align}
\label{CS23}
\frac{dQdp}{\hbar} = \e^{-M(q,p)} \frac{dqdp}{\hbar}\, .
\end{align}
Note, however, the ranges of the integrations for the canonical variables and possibly the boundary conditions for the canonical variables,
might be changed by the redefinition in (\ref{CS21}) and may give a non-trivial result. 

When $N\geq 2$, however, the situation could be drastically changed. 
Instead of (\ref{CS20}), one may consider the following commutation relation, 
\begin{align}
\label{CS24}
\left[ q^i, p_i \right] = i\hbar \e^{M^i_{\ j}(q^k,p_k)}\, ,
\end{align}
which gives the metric in the phase space, 
\begin{align}
\label{CS25}
ds^2 = \sum_{i,j=1}^N g^j_{\ i}dq^i dp_j \, , \quad g^j_{\ i} \equiv \left( L^{-1} \right)^j_{\ i} \, , L^i_{\ j}\equiv \e^{M^i_{\ j}(q^k,p_k)} \, .
\end{align}
Here $\left( L^{-1} \right)^j_{\ i}$ is the inverse matrix of $L^i_{\ j}$ when we regard $L^i_{\ j}$ as $N\times N$ matrix, 
$\sum_{k=1}^N L^i_{\ k} \left( L^{-1} \right)^k_{\ j} = \delta^i_{\ j}$. 
In fact, the metric $g^i_{\ i}$ gives the following volume form
\begin{align}
\label{CS26}
dV = \det \left( g^j_{\ i} \right) \prod_{i=1}^N \left( dq^i dp_i \right)\, .
\end{align}
We should note that due to the symplectic structure of the phase space, $\det \left( g^j_{\ i} \right)$ is a Pfaffian. 
In the case of $\e^{M^i_{\ j}(q^k,p_k)}=\e^{\frac{1}{N}M\left( q^i, p_i \right)}$, $dV$ becomes the previous expression of the 
general measure multiplied by $\hbar^N$, 
\begin{align}
\label{CS26}
dV = \hbar^N \e^{-M\left(q^i,p_i\right)}\prod_{i=1}^N \left( \frac{dq^i dp_i}{\hbar} \right) \, .
\end{align}
Different from the case that $N=1$, the metric in (\ref{CS25}) and therefore the commutation relations in (\ref{CS24}) cannot be rewritten 
in a diagonal form like $\left[ Q^i, P_i \right] = i\hbar \delta^i_{\ j}$ by the redefinition of the variables $Q^i=Q^i\left( q^j, p_j \right)$, 
$P_i =P_i \left( q^j, p_j \right)$ if the curvatures given by the metric in (\ref{CS25}) do not vanish. 

For example, in the case of the three-parameter entropy (\ref{general6}), by using (\ref{CS11}) with (\ref{CS18}), 
Eq.~(\ref{CS24}) has the following form, 
\begin{align}
\label{CS27}
\left[ q^i, p_j \right] = \frac{i\hbar \e^{\frac{1}{\gamma} \left[ \left( 1 + \frac{\alpha { c^3} \pi \left( x^2 + y^2 + z^2 \right)}{{ \hbar} G \beta} \right)^\beta - 1 \right] }}
{{ 4\pi R^2} \delta\left( R^2 - x^2 - y^2 -z^2 \right)} \delta^i_{\ j} \, .
\end{align}
Of course, the inverse power of the delta function does not have a physical meaning. 
We should note, however, that delta function $\delta(x)$ can be defined by, 
\begin{align}
\label{CS28}
\delta (x) \equiv \lim_{\lambda\to \infty} \sqrt{\frac{\lambda}{\pi}} \e^{-\lambda x^2}\, .
\end{align}
Therefore instead of (\ref{CS27}), by using sufficiently large parameter $\lambda$, one may consider the following commutation relation, 
\begin{align}
\label{CS29}
\left[ q^i, p_j \right] = \frac{i\hbar}{ 4\pi R^2} \sqrt{\frac{\pi}{\lambda}} \e^{\frac{1}{\gamma} \left[ \left( 1 + \frac{\alpha { c^3} 
\pi \left( x^2 + y^2 + z^2 \right)}{{ \hbar}G \beta} \right)^\beta - 1 \right] 
+ \lambda \left( R^2 - x^2 - y^2 -z^2 \right)^2 } \delta^i_{\ j} \, , \quad i,j=x,y,z, \quad \left(q^x, q^y, q^z \right)=\left(x,y,z \right)\, ,
\end{align}
which might be the origin of the three-parameter entropy (\ref{general6}). 

In the commutation relations (\ref{CS24}), the quantity $\e^{M^i_{\ j}(q^k,p_k)}$ in the r.h.s. is an operator and therefore we need to specify the ordering 
of $q^k$ and $p_k$ so that the quantity $\e^{M^i_{\ j}(q^k,p_k)}$ becomes a well-defined operator. 
After defining the ordering, the commutators of the operator $\e^{M^i_{\ j}(q^k,p_k)}$ generate an infinite number of operators. 
In fact $\left[ \e^{M^i_{\ j}(q^k,p_k)}, p_i \right]= i\hbar \frac{\partial \e^{M^i_{\ j}(q^k,p_k)}}{\partial q^i}$ and the commutators including 
$\frac{\partial \e^{M^i_{\ j}(q^k,p_k)}}{\partial q^i}$ give further new operators. 
This structure is very similar to Zamolodchikov's W-algebra \cite{Zamolodchikov:1985wn}, which appeared in some conformal field theories. 

On the other hand, the generalised uncertainty principle based on the introduction of the minimal length 
\cite{Yoneya:1989ai} could generate the modification of the canonical commutation relations. 
The concept of the minimal length is motivated by string theory where the fundamental string has a minimum size. 

As well-known, the conformal field theories also appear in the string theory, which may give fundamental quantum gravity. 
Hence one can conjecture that generalised commutation relations (\ref{CS24}) might have the origin from string theory and/or 
quantum gravity. 

\section{Summary}\label{Sec4}

We tried to construct the microscopic description of generalised entropy, which has been proposed in several recent works 
as the most general entropy construct giving all known physical/statistical entropies as particular representatives. 
Our consideration is based on the micro-canonical formulation and the canonical formulation of thermodynamics. 
It has been shown that the change of the volume in phase space of the system brings us to generalised entropy in the micro-canonical description. 
On the other hand, in the canonical description, we found that the modification of the integration measure in the phase space may
generate different entropies including a few parameter-dependent generalised entropy. In the same way, the thermodynamical foundation 
for an arbitrary finite or infinite number of parameters dependent generalised entropy may be developed.
Finally, some speculative remark is in order.
If quantum gravity has a thermodynamical description, the modification of the corresponding microscopic structures as in this paper 
might clarify its basic formulation. 

\section*{Acknowledgments}

The authors are indebted to JSPS because this work was completed during the visit of SDO to Nagoya University by 
FY2023 JSPS Invitational Fellowships for Research in Japan (S23013).
This work has been partially supported by MICINN (Spain), project PID2019-104397GB-I00 and 
by the program Unidad de Excelencia Mar{\' i}a de Maeztu CEX2020-001058-M (SDO).

\end{document}